\documentclass[letterpaper]{jpconf}
\usepackage{graphicx}

\def\bea{\begin{eqnarray}}
\def\eea{\end{eqnarray}}

\begin{document}
\title{The equivalence of fluctuation scale dependence and autocorrelations}

\author{Duncan J Prindle and Thomas A Trainor}

\address{CENPA 354290 University of Washington, Seattle WA 98195 USA}

\ead{prindle@npl.washington.edu,trainor@hausdorf.npl.washington.edu}

\begin{abstract}
We define optimal per-particle fluctuation and correlation measures, relate fluctuations and correlations through an integral equation and show how to invert that equation to obtain precise autocorrelations from fluctuation scale dependence. We test the precision of the inversion with Monte Carlo data and compare autocorrelations to conditional distributions conventionally used to study high-$p_t$ jet structure.
\end{abstract}.

\section{Introduction}

Fluctuations in nuclear collisions measured at a single bin size or {\em scale} could arise from many possible configurations of a multiparticle momentum distribution and are therefore difficult to interpret. However, the {\em scale dependence} of fluctuations over a significant scale interval does contain detailed information about multiparticle correlations which can be extracted with the proper techniques~\cite{inverse}. Information about the absolute location of event-wise structure is lost, but those aspects depending only on position {\em difference} are retained in the form of autocorrelation distributions~\cite{auto1}. In this paper we consider fluctuations on binned momentum space, Pearson's correlation coefficient and autocorrelation density ratios for multiplicity $n$ and transverse momentum $p_t$. Autocorrelations can be inferred directly from pair ratios or from a fluctuation/autocorrelation {\em integral equation} which we derive. Inverting the integral equation we obtain autocorrelations which can be compared with more conventional conditional distributions.

\section{Fluctuations and correlations on binned spaces}

Correlation analysis of nuclear collisions reveals information arising from event-wise changes in multiparticle momentum distributions.
The data system is an ensemble of event-wise particle distributions on momentum space $(p_t,\eta,\phi)$ or $(y_t,\eta,\phi)$, where $p_t$ is transverse momentum, $m_t$ is transverse mass, $\eta$ is pseudorapidity, $\phi$ is azimuth and $y_t \equiv \ln\{(m_t + p_t)/m_0\}$ is transverse rapidity with pion mass assigned to $m_0$. The momentum space is bounded by a detector acceptance, and the space within the acceptance is binned according to one or more bin sizes. Particle number $n$ is distributed on the full momentum space. It is useful to think of transverse momentum $p_t$ or rapidity $y_t$ as a continuous measure distributed  on angle subspace $(\eta,\phi)$ and sampled by individual particles. Number $n$ and transverse momentum $p_t$ or rapidity $y_t$ correlations can be considered both separately and in conjunction. 

\begin{figure}[h]
\begin{minipage}{38pc}\hfil
\includegraphics[width=8pc]{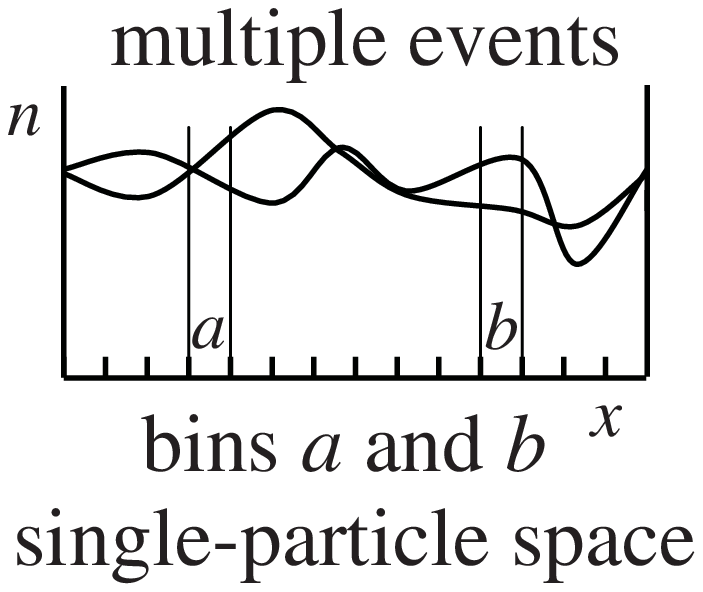} \hfil
\includegraphics[width=8pc]{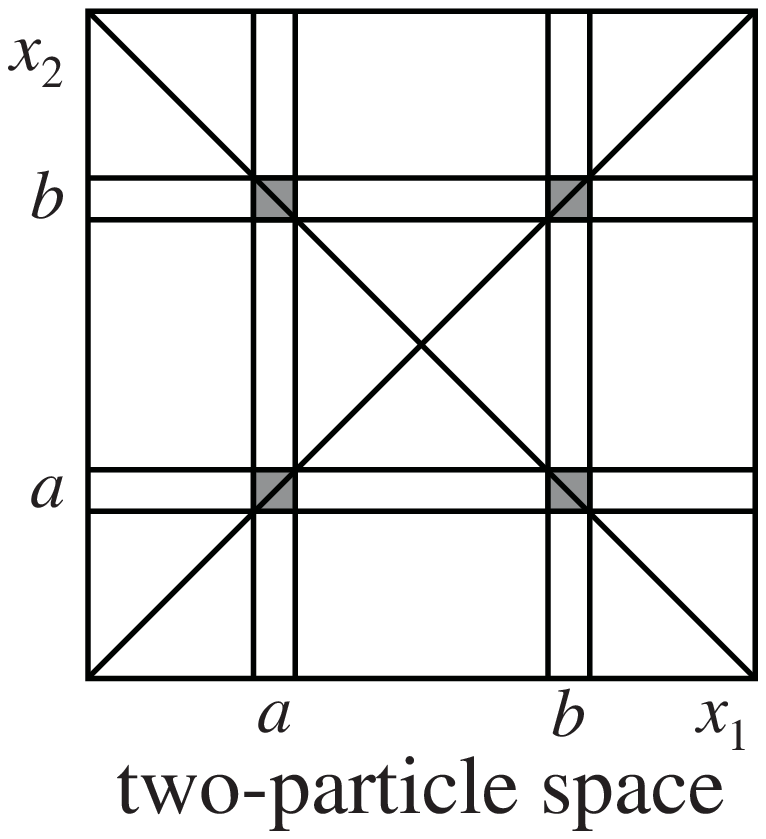} \hfil
\includegraphics[width=11.6pc]{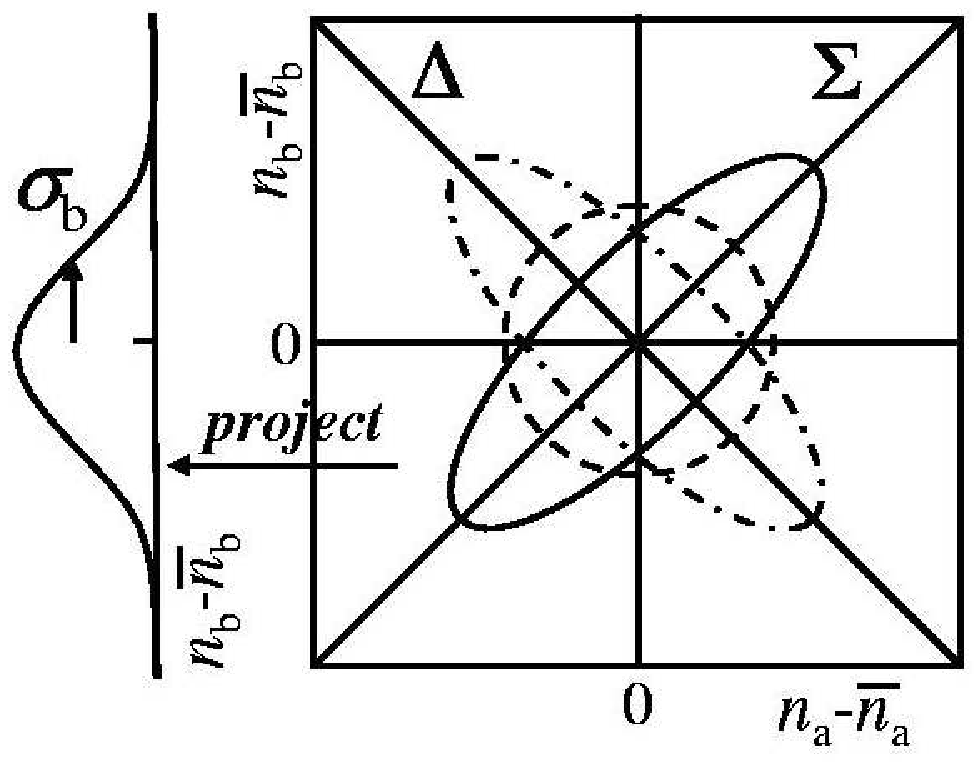} \hfil
\end{minipage} 
\caption{\label{fluctbin}  Event-wise distributions on a binned primary space, corresponding two-particle space and possible distributions of bin contents for selected bins $a$ and $b$ in the primary space.}
\end{figure}

An ensemble of event-wise histograms on momentum subspace $x$ is represented schematically in Fig.~\ref{fluctbin} (first panel), with bins $a,\,b$ singled out. Events can be compared with the ensemble-mean distribution to determine {\em relative} information, measured by fluctuations of bin contents about their means. The ensemble mean of event-wise pair distributions on space $(x_1,x_2)$ in Fig.~\ref{fluctbin} (second panel) can be compared to a reference distribution consisting of cartesian products of single-particle mean distributions. The {\em difference} reveals correlations in the two-particle distribution corresponding to fluctuations in the single-particle distribution~\cite{cltps}. That relation is the basis for the integral equation connecting fluctuations to correlations described below. 


 Fig.~\ref{fluctbin} (third panel) sketches possible frequency distributions on number combinations $(n_a,n_b)$ from bin pair $(a,b)$ in the first panel. The ellipses represent half-maximum contour lines for gaussian-random fluctuations. The three cases correspond to correlation (solid curve), anticorrelation (dash-dot curve) and no correlation (dashed curve) between bins, the last being expected from a mixed-pair or {\em central limit} reference. The 2D frequency distribution is characterized by two marginal variances and a covariance. The marginal variances for bins $a$ and $b$ are given by $\sigma^2_{a,b} = \overline{(n - \bar n)_{a,b}^2} = \overline{n^2_{a,b}} - \bar n^2_{a,b}$. The covariance between those bins is given by $\sigma^2_{ab} = \overline{(n - \bar n)_{a}(n - \bar n)_{b}} = \overline{n_a\, n_b} - \bar n_a\, \bar n_b = \sigma^2_\Sigma - \sigma^2_\Delta$, where $\sigma^2_\Sigma$ and $\sigma^2_\Delta$ are variances along sum and difference diagonals in the third panel.
Pearson's correlation coefficient is a measure of {\em relative} covariance~\cite{pearson}. For bin pair $(a,b)$ it is defined by $r_{ab} \equiv \sigma^2_{ab} / \sqrt{\sigma^2_a\, \sigma^2_b} = \{ \sigma^2_\Sigma - \sigma^2_\Delta\} / \{ \sigma^2_\Sigma + \sigma^2_\Delta\} \in [-1,1]$. The numerator is the ($a,b$) covariance and the denominator is the geometric mean of the marginal $a$ and $b$ variances. That coefficient is our basic correlation measure. Quantities $r_{ab}$, determined for all histogram bin pairs in the second panel, completely represent fluctuations on space $x$. Variances and covariances depend on the bin size or scale on $x$. The {\em scale dependence} of fluctuations is in turn directly related to two-particle correlations, as described in this paper.



\section{Object and reference distributions} \label{objref}

We distinguish between an object distribution, part of whose correlation content we wish to measure, and a reference distribution which contains by construction information in the object distribution we wish to ignore. We determine object and reference distributions for number $n$ correlations and $p_t$ correlations (the latter require a slightly different treatment, as described below).
One single-particle reference for fluctuation measurements is the ensemble-mean histogram on $x$. Variances and covariances for single bins and bin pairs measure the average information in the event-wise object distribution relative to the ensemble mean. 


For two-particle distributions on $(x_1,x_2)$, object pair density $\rho_{obj}(x_1,x_2)$ is constructed from {\em sibling} pairs taken from same events, and reference pair density $\rho_{ref}(x_1,x_2)$ is constructed from {\em mixed} pairs taken from different but similar events. The reference could also be a Cartesian product of single-particle means (therefore {\em factorizable} by construction). 
Object and reference distributions are combined in several ways:
1) density of correlated pairs: $\Delta \rho(x_1,x_2) \equiv \rho_{obj}(x_1,x_2) - \rho_{ref}(x_1,x_2)$; 2) density of correlated pairs {\em per particle pair}:  $\Delta \rho(x_1,x_2) / \rho_{ref}(x_1,x_2) \equiv \rho_{obj}(x_1,x_2) / \rho_{ref}(x_1,x_2) - 1$; 3) density of correlated pairs {\em per particle}:  $\Delta \rho(x_1,x_2) / \sqrt{\rho_{ref}(x_1,x_2)} \equiv \{\rho_{obj}(x_1,x_2) - \rho_{ref}(x_1,x_2) \} / \sqrt{\rho_{ref}(x_1,x_2)}$. 1) and 2) are conventional correlation measures. 3) is unconventional, but closely related to Pearson's correlation coefficient described in the previous section: a relative covariance measure which does not depend on the absolute number of particles {\em per se}. Pearson's coefficient is ideally suited for testing linear superposition in the context of heavy ion collisions.

\section{Autocorrelations}

\begin{figure}[h]
\begin{minipage}{19pc} \vspace{.19in}
\includegraphics[width=19pc,height=9.6pc]{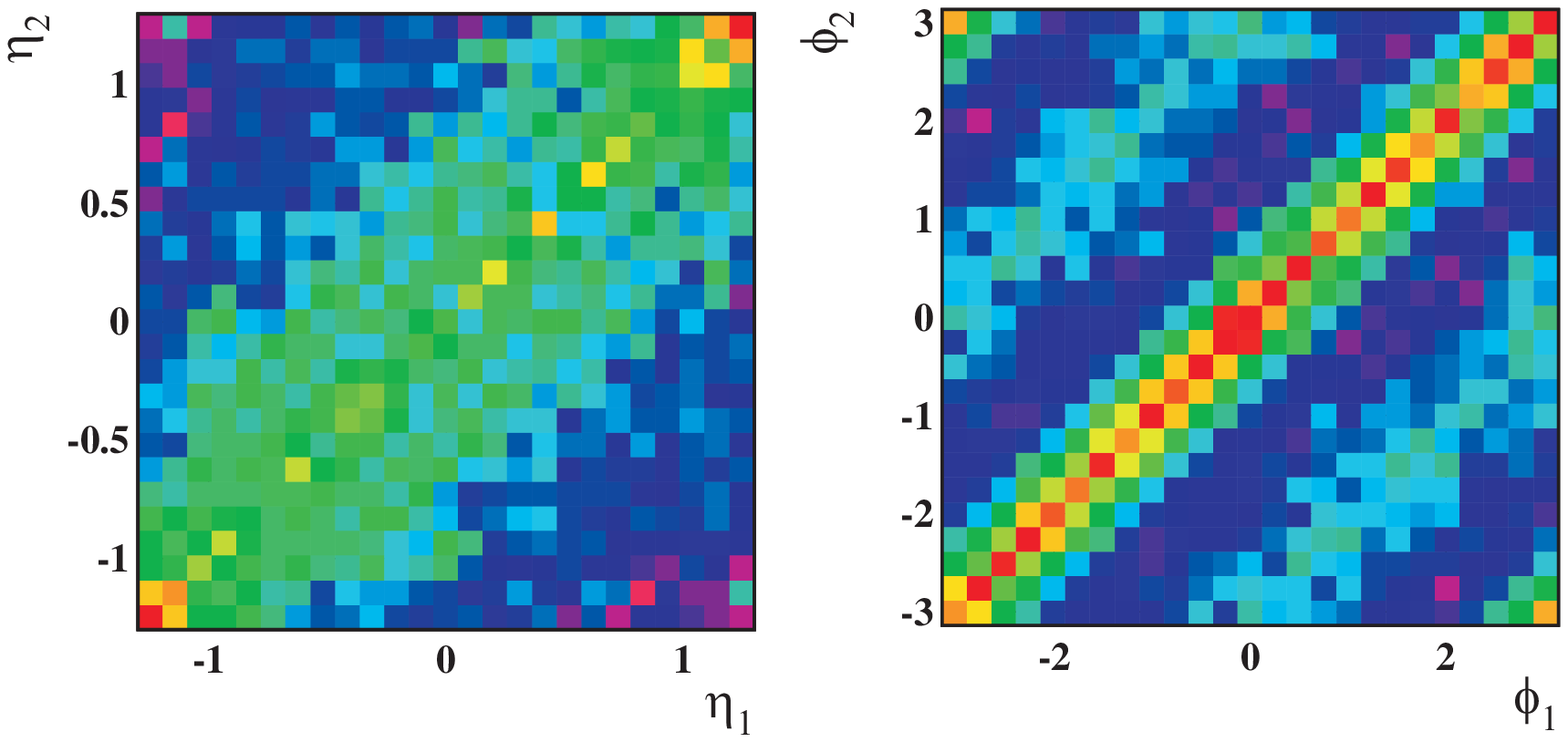}
\end{minipage} 
\begin{minipage}{19pc}
\includegraphics[width=9pc]{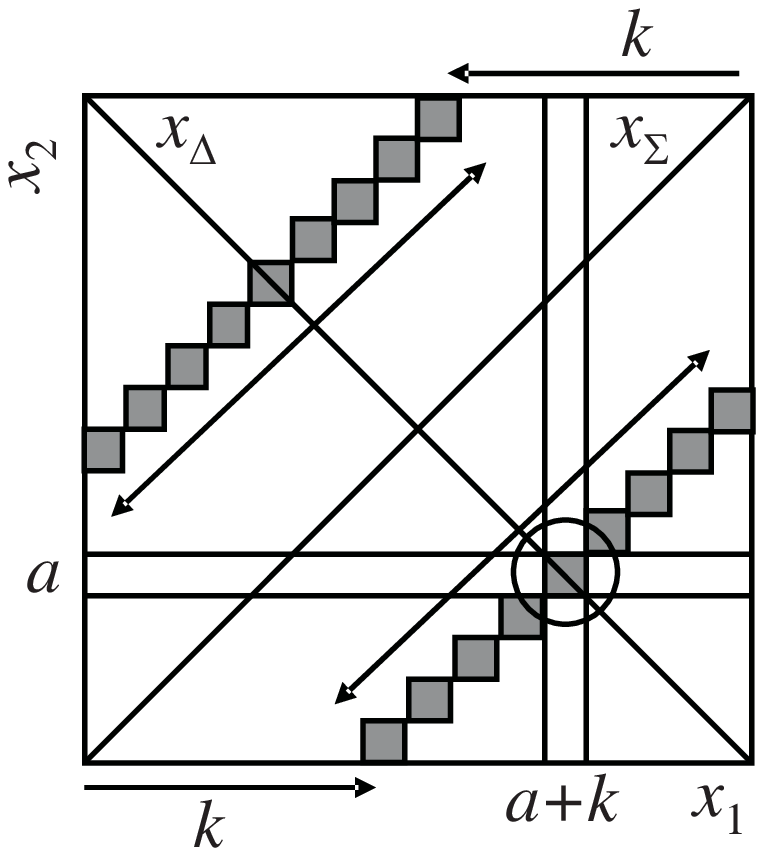}
\includegraphics[width=9pc]{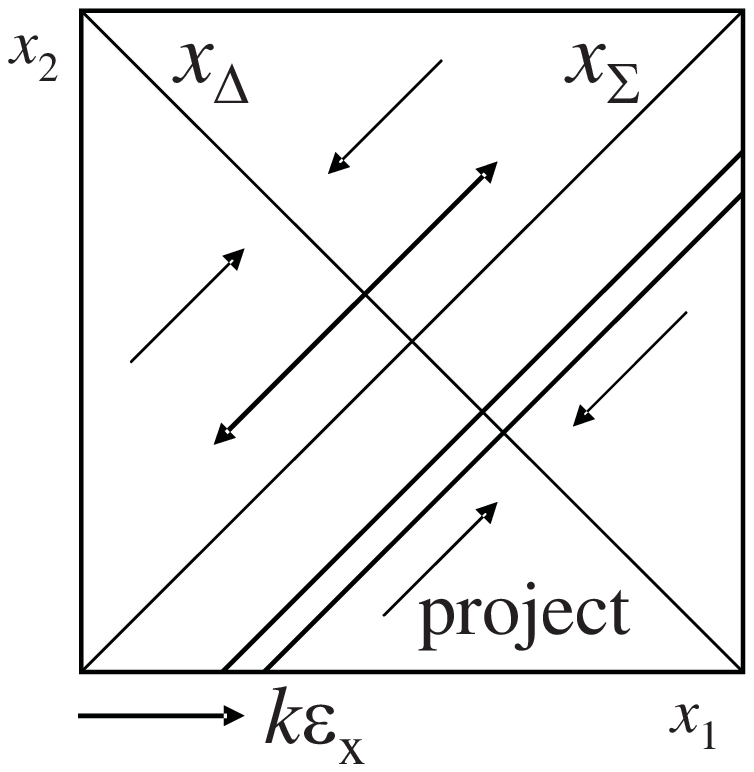}
\end{minipage} 
\caption{\label{autocorr}  Two-particle correlations on $\eta$ and $\phi$ for central Au-Au collisions at 130 GeV, schematic of a binned two-particle space illustrating an autocorrelation average along the $k^{th}$ diagonal and a similar averaging procedure performed directly on difference variable $x_\Delta$.}
\end{figure}

The autocorrelation concept was first introduced to time-series analysis in the form $A(\tau) \equiv 1/T \int_{-T/2}^{T/2} f(t)\, f(t + \tau)\, dt$, where $\tau$ is the {\em lag}~\cite{auto1}. The concept is most useful when function $f(t)$ is {\em stationary}: its correlation structure does not depend on absolute location on time. The information in $f(t)$ is then fully represented by the autocorrelation distribution on $\tau$. The autocorrelation concept can be generalized to spatial correlations. If event-wise structure is randomly positioned on space $x$ then the corresponding ensemble-average two-point distribution on $(x_1,x_2)$ is stationary (not depending on absolute position on $x$). Distributions in Fig.~\ref{autocorr} (left two panels) of measure $\Delta \rho(x_1,x_2) / \rho_{ref}(x_1,x_2)$ on $\eta$ and $\phi$ are typical of Au-Au collisions at RHIC~\cite{axialcd}. Two-particle distributions can also be defined on sum and difference variables: $\rho(x_1,x_2) \rightarrow \rho(x_\Sigma,x_\Delta)$, with $x_1, x_2 \rightarrow x_\Sigma \equiv x_1 + x_2, x_\Delta \equiv x_1 - x_2$. The data distributions in Fig.~\ref{autocorr} exhibit stationarity---they do not depend on sum variable $x_\Sigma$---in which case we have $\rho(x_\Sigma,x_\Delta) \rightarrow \rho(x_\Delta)$ to good approximation. We then {\em average} the two-particle density over $x_\Sigma$ to obtain the autocorrelation density on $x_\Delta$ (still a 2D density, not a projection).

The autocorrelation on difference variables can be constructed in two ways as shown in Fig.~\ref{autocorr} (last two panels): 1) bin space $x$ with {\em microbins} of size $\epsilon_x$ and average the bin contents of space $(x_1,x_2)$ along diagonals, as discussed further below, 2) bin difference variable $x_\Delta$ in space $(x_1,x_2)$ directly and form the corresponding pair histogram~\cite{inverse}.  In both cases, care must be taken to insure that true averages are obtained and not projections by integration. We define autocorrelation densities $\rho(x_\Delta)$, autocorrelation histograms $A_k(\epsilon_x) \simeq \epsilon^2_x \, \rho(2k\epsilon_x)$ and {\em joint} autocorrelations on two or more difference variables. If the primary distribution on $x$ is truly stationary, the autocorrelation is a {\em lossless compression} of the two-particle momentum distribution to a lower-dimensional space.

\section{Autocorrelation density ratio for number correlations}

We now define a universal correlation measure for nuclear collisions.
Returning to Pearson's correlation coefficient we make the following approximation
\bea \label{pears}
r_{ab} \equiv \frac{\sigma^2_{ab}}{\sqrt{\sigma^2_{a}\, \sigma^2_{b}}} = \frac{\overline{(n - \bar n)_a(n - \bar n)_b}}{{\sqrt{\overline{(n - \bar n)^2_a}\,\, \overline{(n - \bar n)^2_b}}}} \rightarrow \frac{\overline{(n - \bar n)_a(n - \bar n)_b}}{{\sqrt{\bar n_a\,  \bar n_b}}} = \frac{\overline{n_a\, n_b} - \bar n_a\,  \bar n_b}{{\sqrt{\bar n_a\,  \bar n_b}}}.
\eea 
We replace the marginal number variances in the denominator of $r_{ab}$ by their Poisson values. The result is the histogram equivalent of density ratio $\Delta \rho(x_1,x_2) / \sqrt{\rho_{ref}(x_1,x_2)} $ previously defined. This relative covariance can be interpreted as the number of (anti)correlated pairs per particle (explicit in the last expression). The third combination of object and reference distributions in Sec.~\ref{objref} and the modified Pearson's normalized covariance in Eq.~(\ref{pears}) are equivalent. This density ratio is the basic correlation measure for any bin pair $(a,b)$ on space $x$.
We then define an autocorrelation in terms of density ratios for sets of bins $(a,b)$ on $(x_1,x_2)$: we combine density ratios as averages along diagonals $k$, assuming the basic distribution on $(x_1,x_2)$ is approximately stationary. The average on index $a$ along the $k^{th}$ diagonal in Fig.~\ref{autocorr} (third panel) is
\bea \label{eq2}
\frac{\Delta A_k(n)}{\sqrt{A_{k,ref}(n)}} \equiv \left\{ {\frac{\overline{(n - \bar n)_a(n - \bar n)_{a+k}}}{{\sqrt{\bar n_a\,  \bar n_{a+k}}}}} \right\}_{\bar a} \equiv \epsilon_{x_\Delta}\, \frac{\Delta \rho(n;k \epsilon_{x_\Delta}) }{ \sqrt{\rho_{ref}(n;k \epsilon_{x_\Delta})} }.
\eea
This is the autocorrelation definition for analysis of nuclear collisions on angle space $(\eta,\phi)$. This density ratio is an intensive correlation measure which precisely measures relative correlations even for excursions of object and reference densities over orders of magnitude.

\section{Extension to transverse momentum $p_t$ correlations}


We now extend the definition of the density ratio and its autocorrelation to distributions of transverse momentum $p_t$ on $(\eta,\phi)$. Measurement of $\langle p_t \rangle$ fluctuations is described in~\cite{ptprl}. We treat $p_t$ as a continuous measure distributed on space $x$, with scalar sums of particle $p_t$ in histogram bins. We could write the Pearson's coefficient for transverse momentum by analogy with number correlations in terms of difference $(p_t - \bar p_t)$. However, the corresponding per-particle $p_t$ variance can be expressed as the sum of three terms:
$\overline{(p_t - \bar p_t)^2}/\bar n = \overline{(p_t -  n\, \hat p_t)^2}/\bar n + 2\hat p_t\,\overline{(p_t -  n\, \hat p_t)(n - \bar n)}/\bar n + \hat p^2_t\, \overline{(n - \bar n)^2}/\bar n$, a `$\langle p_t \rangle$' variance, a $p_t$-$n$ covariance and a number variance. The three terms have forms similar to Pearson's normalized covariance, but with $a = b$ defining corresponding {\em normalized variances} (also called `scaled variances'). Each term is important in the overall problem of particle and $p_t$ production.  We therefore want to describe the structure of event-wise $p_t$ distributions in terms of $(p_t - n\, \hat p_t)$, independent of but coordinated with the structure of number distributions described in terms of $(n - \bar n)$.
Pearson's correlation coefficient for {\em transverse momentum} correlations thus takes the form
\bea \label{ptrab}
r_{ab} \sim \frac{\overline{(n - \bar n)_a(n - \bar n)_b}}{{\sqrt{\bar n_a\,  \bar n_b}}} \rightarrow \frac{\overline{(p_t - n\, \hat p_t)_a(p_t - n\, \hat p_t)_b}}{{\sigma^2_{\hat p_t}\sqrt{\bar n_a\,  \bar n_b}}},
\eea 
{\em i.e.,} a relative $p_t$ covariance as opposed to a relative number covariance. The geometric mean of marginal variances in the denominator is in this case replaced by $\sigma^2_{\hat p_t}\sqrt{\bar n_a\,  \bar n_b}$, the mean of central-limit expectations for the $p_t$ variances. The factor $\sigma^2_{\hat p_t}$ is however omitted in what follows to be consistent with the first term of the per-particle $p_t$ variance expansion above Eq.~(\ref{ptrab}). Factors $\sigma^2_{\hat p_t}$ or $\hat p_t^2$ may be introduced as necessary in a subsequent interpretation stage. The corresponding $p_t$ density-ratio autocorrelation is
\bea
\frac{\Delta A_{kl}(p_t:n)}{\sqrt{A_{kl,ref}(n)}} \equiv \left\{  {\frac{\overline{(p_t - n\, \hat p_t)_{ab}(p_t - n\, \hat p_t)_{a+k,b+l}}}{{\sqrt{\bar n_{ab}\,  \bar n_{a+k,b+l}}}}} \right\}_{\overline{ab}} \equiv \epsilon_{\eta_\Delta}\epsilon_{\phi_\Delta}\, \frac{\Delta \rho(p_t:n;k \epsilon_{\eta_\Delta}, l \epsilon_{\phi_\Delta}) }{ \sqrt{\rho_{ref}(n;k \epsilon_{\eta_\Delta}, l \epsilon_{\phi_\Delta})} }.
\eea
That expression is formulated explicitly in terms of 2D correlations on $(\eta,\phi)$. An analogous 2D expression can be derived for number correlations from Eq.~(\ref{eq2}).

\begin{figure}[h]
\begin{minipage}{19pc}
\includegraphics[width=9pc]{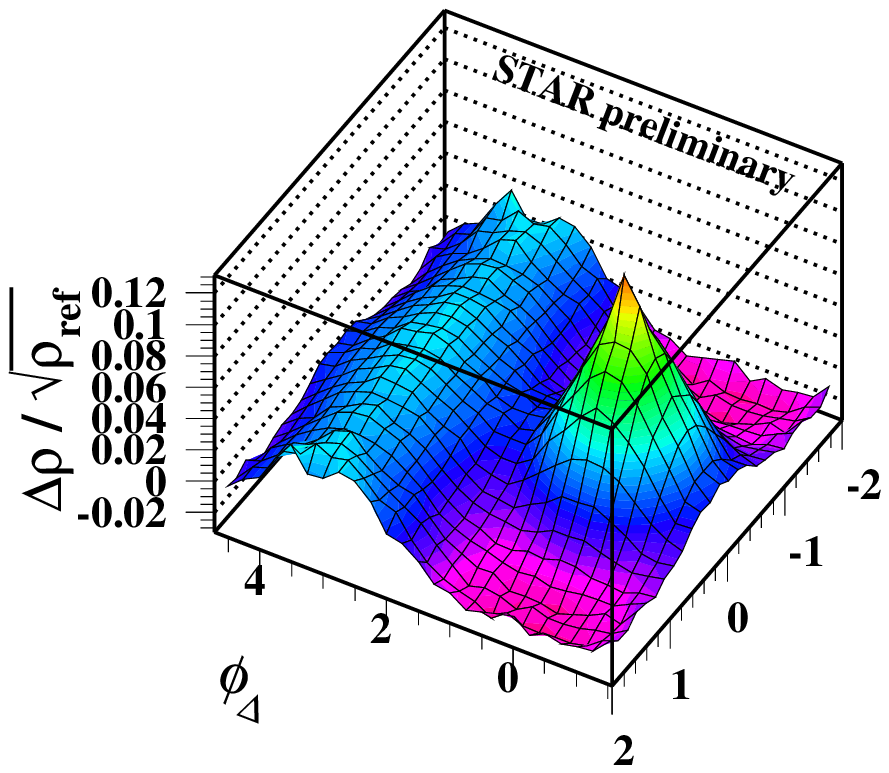}
\includegraphics[width=9pc]{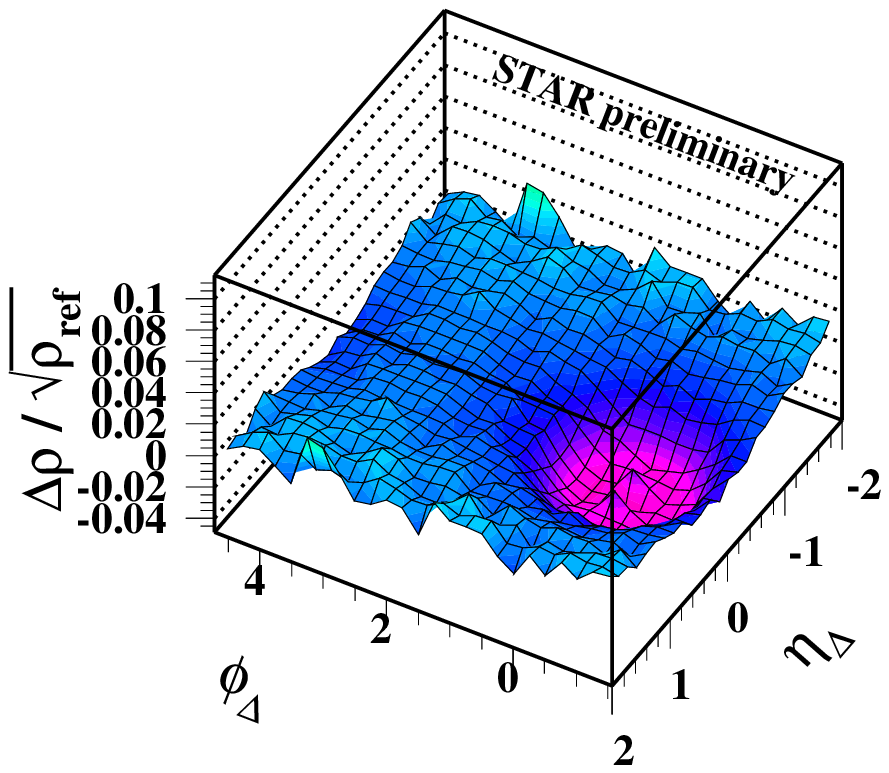}
\end{minipage} 
\begin{minipage}{19pc}
\includegraphics[width=9pc]{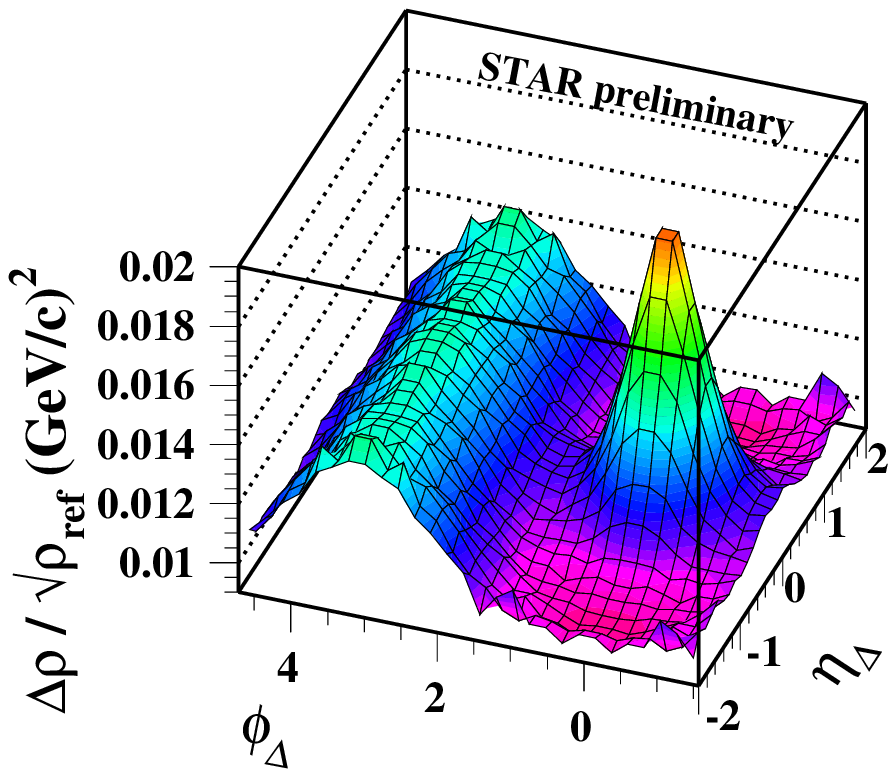}
\includegraphics[width=9pc]{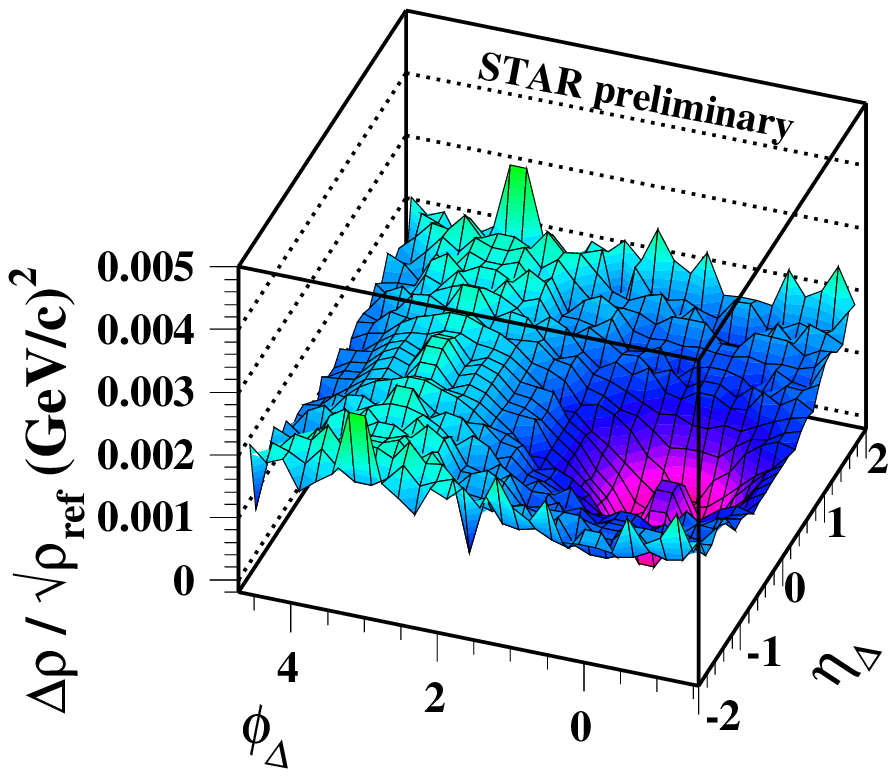}
\end{minipage} 
\caption{\label{directauto}  Autocorrelations for 200 GeV p-p collisions determined by direct pair counting: number correlations (left panels) and $p_t$ correlations (right panels) and for charge-independent (left) and  charge-dependent or net-charge (right) correlations in each case.}
\end{figure}  

Fig.~\ref{directauto} shows joint autocorrelations of per-particle density ratios on angle difference variables 
for number correlations (first two panels) and $p_t$ correlations (second two panels) and for charge-independent (like-sign $+$ unlike-sign) pair combinations (left panel of each pair) and charge-dependent (like-sign $-$ unlike-sign) pair combinations (right panel of each pair)~\cite{jeffismd}. Contributions from self pairs were excluded. These joint autocorrelations, determined directly by pair counting, provide full access to all angular correlations on $(\eta,\phi)$ but are computationally expensive for larger event multiplicities. We can also obtain these autocorrelations by inverting fluctuation scale dependence, but with much less computation effort.

\section{Relating fluctuations and correlations}

Fluctuation scale dependence results from event-wise correlation structure in single-particle distributions which can be extracted by solving an integral equation. We first define differential scale-dependent fluctuation measures based on variance differences for number $n$ and transverse momentum $p_t$ fluctuations, then derive the integral equation connecting fluctuations and correlations.
%
Number variance difference $\Delta \sigma^2_{n/}(\delta x) \equiv \overline{(n(\delta x) - \bar n(\delta x))^2}/\bar n(\delta x) - 1$ compares the normalized number variance at bin scale $\delta x$ to the central-limit or Poisson expectation 1 for that quantity. The difference reflects correlations in the number distribution beyond that of a random distribution of points. The variance difference can be interpreted as the total number of correlated pairs (per-bin number variance $\sigma_n^2$) minus the number of Poisson-correlated pairs (self pairs $\bar n$) per particle (divided by $\bar n$). The corresponding variance difference for $p_t$ fluctuations is $\Delta \sigma^2_{p_t:n}(\delta x) \equiv \overline{(p_t(\delta x) -  n(\delta x)\, \hat p_t)^2}/\bar n(\delta x) - \sigma^2_{\hat p_t}$. Those per-particle quantities, defined in terms of normalized variances, are consistent in form with Pearson's normalized covariance.



\begin{figure}[h]
\begin{minipage}{39pc}
\includegraphics[width=13pc,height=11pc]{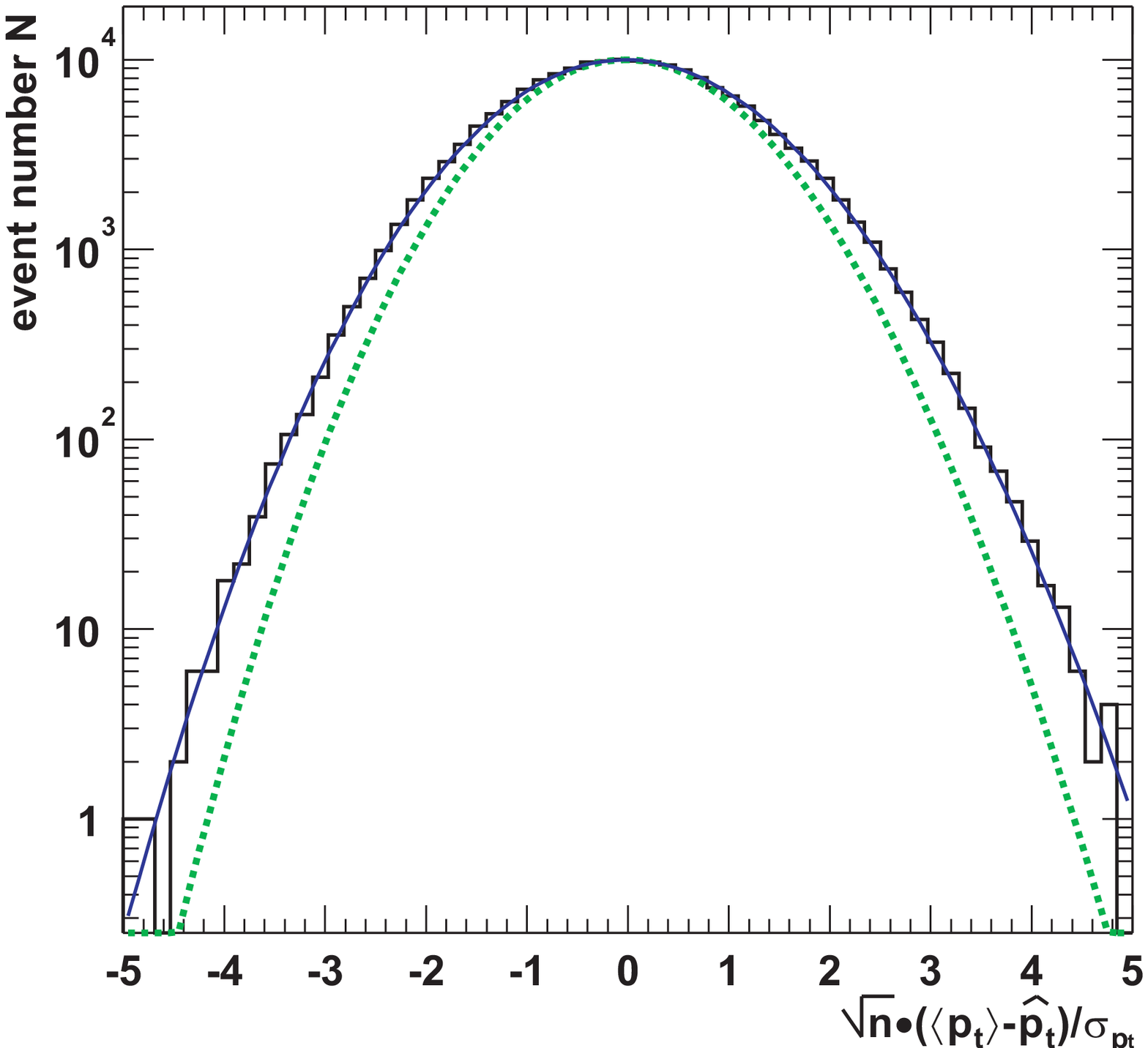}
\includegraphics[width=13pc]{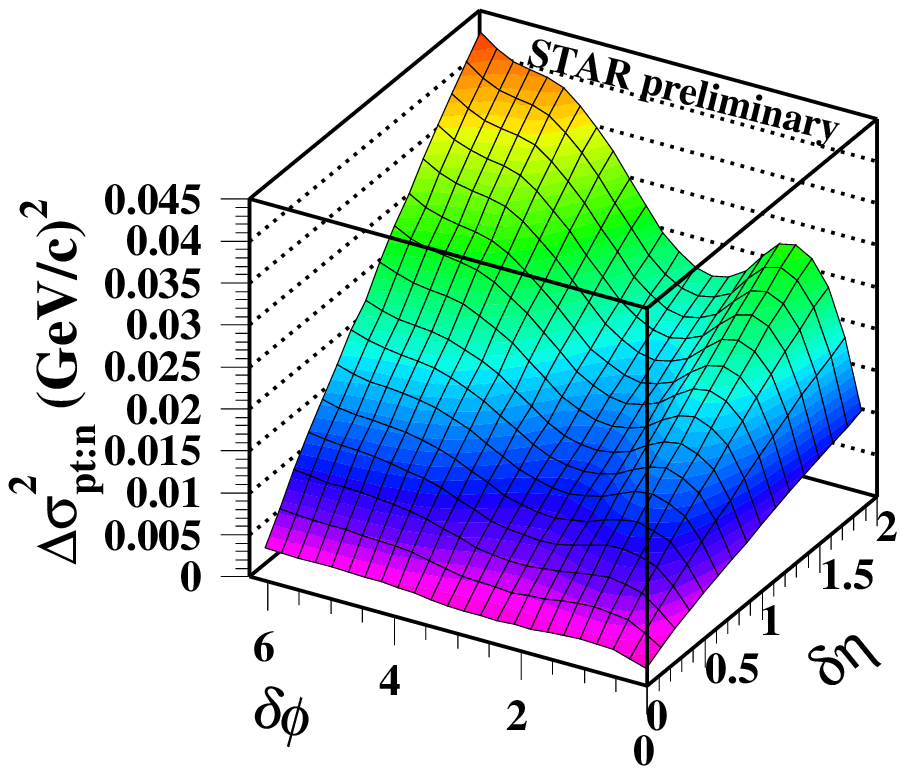}
\includegraphics[width=13pc]{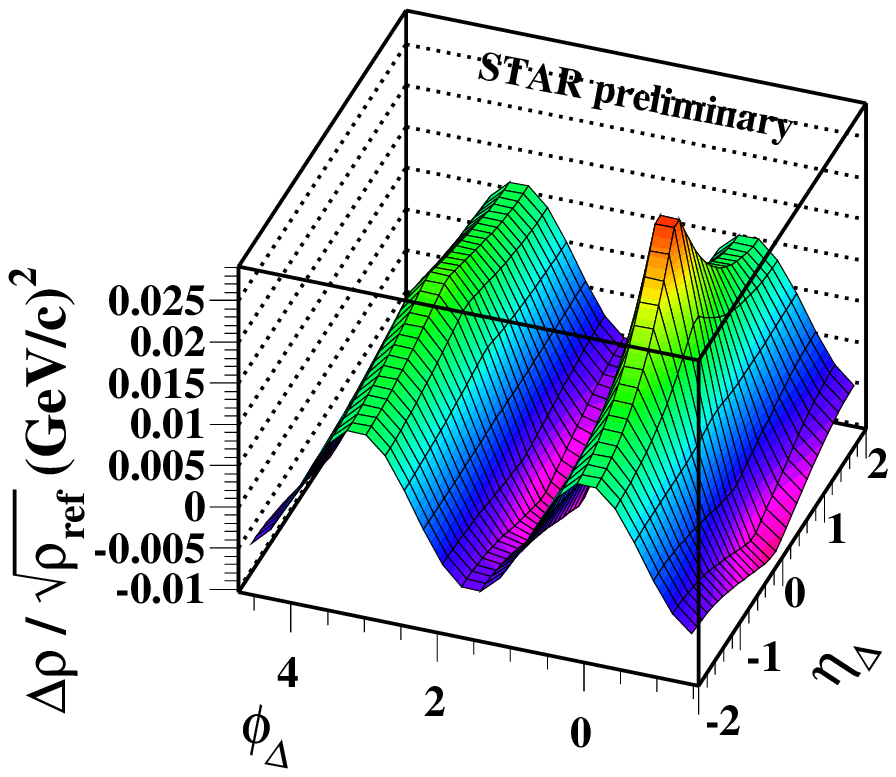}
\end{minipage} 
\caption{\label{fluctcorr}  $\langle p_t \rangle$ fluctuations measured at the STAR detector acceptance (histogram) compared to a central-limit reference (dotted curve), the scale dependence of $\langle p_t \rangle$ fluctuations within the STAR acceptance and the corresponding $p_t$ autocorrelation obtained by inversion.}
\end{figure}

A fluctuation measurement at a single scale (STAR detector acceptance) is shown in the first panel of Fig.~\ref{fluctcorr}~\cite{ptprl}. The frequency histogram on random variable $(p_t - n\, \hat p_t)/(\sqrt{\bar n}\, \sigma_{\hat p_t})$ is compared to a central-limit reference (green dotted curve, $\sigma_{\hat p_t}$ is the ensemble-average single-particle variance). The $p_t$ variance difference $\Delta \sigma^2_{p_t:n}(\delta x)$ defined previously compares variances of data and reference to reveal a variance excess. Questions then arise how to interpret the fluctuation measurement and how to compare it to measurements made with other detectors.

Fluctuation measurements in different bin-size or scale intervals determine different regions on a {\em common} distribution representing fluctuation scale dependence, as shown in the second panel of Fig.~\ref{fluctcorr}~\cite{ptsca}. The variance difference from the first panel corresponds to the single point at the apex of the surface in the second panel. The surface is obviously structured, but what does the structure mean? Fluctuation scale dependence is the running integral of an autocorrelation. The corresponding integral equation is a linear relation between a variance difference and an autocorrelation, including a kernel representing the binning scheme. We can express the per-particle variance difference on scales $(\delta \eta,\delta \phi)$ as 2D discrete integral
\bea \label{inv}
\Delta \sigma^2_{p_t:n}(m \, \epsilon_\eta, n \, \epsilon_\phi) 
= 4  \sum_{k,l=1}^{m,n} \epsilon_\eta \epsilon_\phi & &\hspace{-.25in} K_{mn;kl}  \,   \frac{\Delta \rho(p_t:n;k\,\epsilon_\eta, l\, \epsilon_\phi) }{ \sqrt{\rho_{ref}(n;k\,\epsilon_\eta, l\, \epsilon_\phi)}} ,
\end{eqnarray}
with kernel $K_{mn;kl} \equiv (m - {k + 1/2})/{m} \cdot (n-{l+1/2})/{n}$ representing the 2D macrobin system. This is a Fredholm integral equation which can be inverted by standard numerical methods to obtain autocorrelation density ratio ${\Delta \rho_{kl}}/{ \sqrt{\rho_{ref,kl}}}$ as a per-particle correlation measure on difference variables $(\eta_\Delta,\phi_\Delta)$~\cite{invtheory}. The third panel of Fig.~\ref{fluctcorr} shows the autocorrelation corresponding to the data in the second panel, with directly interpretable structure~\cite{ptsca}.

\section{Derivation of the integral equation}

\begin{figure}[h]
\begin{minipage}{38pc} \hfil
\includegraphics[width=9pc]{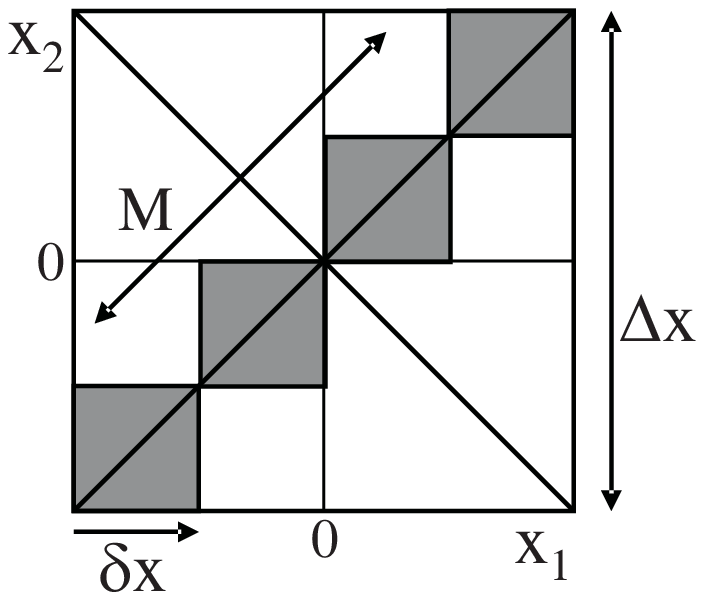} \hfil
\includegraphics[width=9.5pc]{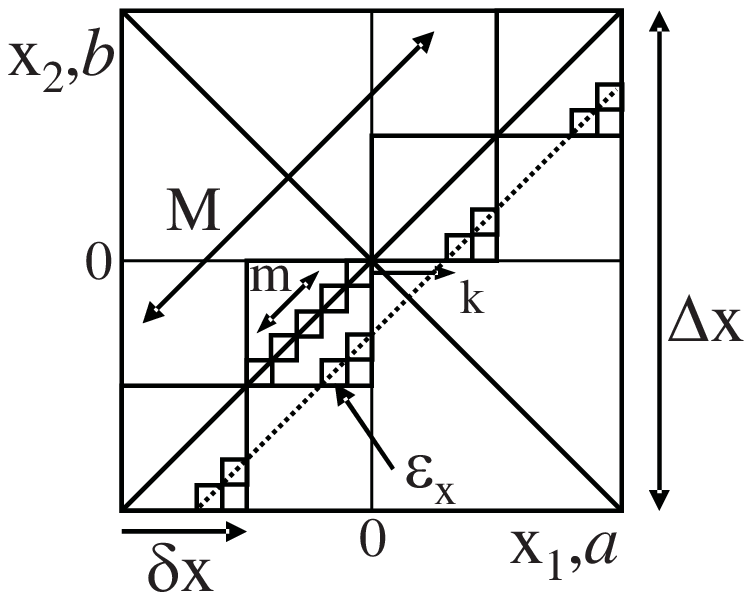} \hfil
\includegraphics[width=18pc]{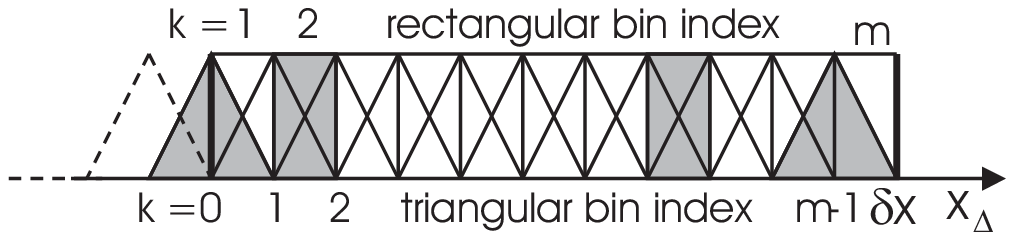} \hfil
\end{minipage} 
\caption{\label{intderiv}  Macrobins of scale $\delta x$ on a two-particle space after binning a primary space $x$, microbins of size $\epsilon_x$ relative to macrobins and two microbin schemes, one on $x$, the other on $x_\Delta$.}
\end{figure}

The derivation relies on coordinating the contents of macrobins, which determine the binning scale for fluctuations, and microbins, which are the basis for the discrete numerical integration of the autocorrelation. We divide the acceptance into macrobins of varying size $\delta x$ (scale dependence) and microbins of fixed size $\epsilon_x$. The scale-dependent variance is an average over all bins of a particular scale within the acceptance as in Fig.~\ref{intderiv} (first panel), where the acceptance is $\Delta x$, the macrobin size is $\delta x$ and the macrobin number is $M$. The average over macrobins is re-expressed as an average over microbins, as shown in Fig.~\ref{intderiv} (second panel). The average over microbins is re-arrange into a sum (over $k$) of different diagonals and an average over microbins on the $k^{th}$ diagonal. The diagonal average is the $k^{th}$ element of an autocorrelation histogram.

The $p_t$ variance difference at scale $\delta x$ in the first line of Eq.~(\ref{inteq}) is re-expressed as a 2D sum over microbins in the second line, with $m$ microbins in each macrobin. Mean multiplicity $\bar n(\delta x)$ in a macrobin relates to the mean microbin multiplicity in that macrobin as $\bar n(\delta x) = m\, \bar n(\epsilon_x)$, which relation is applied in the second line. The single-particle variance in the first line corresponds to self pairs which are excluded from subsequent pair sums. The single-particle variance term is consequently dropped. The third line is a rearrangement of the sum over $(a,b)$ into a sum over diagonal index $k$ and a sum over indices $a, b$ subject to the diagonal constraint $a - b = k$,
\bea \label{inteq}
\Delta \sigma^2_{p_t:n}(\delta x) &\equiv& \overline{(p_t(\delta x) -  n(\delta x)\, \hat p_t)^2}/\bar n(\delta x) - \sigma^2_{\hat p_t} \\ \nonumber
 &=& \sum_{a,b=1}^m \frac{\overline{\{p_t(\epsilon_x) - n(\epsilon_x)\, \hat p_t\}_a\{p_t(\epsilon_x) - n(\epsilon_x)\, \hat p_t\}_b}}{{m\, \bar n(\epsilon_x)}} \\ \nonumber
&=& \sum_{k=1-m}^{m-1} K_{m:k}\, \frac{\bar n_k}{\bar n}\left[\frac{1}{m - |k|} \sum_{1 \leq a,b \leq m}^{a-b=k} \frac{\sqrt{\bar n_a \bar n_b}}{\bar n_k}\cdot \frac{\overline{\{\cdots\}_a\{\cdots\}_b}}{{\sqrt{\bar n_a \bar n_b}}} \right] \\ \nonumber
&\equiv& \sum_{k=1-m}^{m-1} K_{m:k}\, \frac{\Delta A_k(p_t:n;\epsilon_x)}{\sqrt{A_{k,ref}(n;\epsilon_x)}} \\ 
&\rightarrow& 2\sum_{k=1}^{m} K_{m:k}\, \frac{\Delta A_k(p_t:n;\epsilon_x)}{\sqrt{A_{k,ref}(n;\epsilon_x)}} \equiv 2\sum_{k=1}^{m} \epsilon_x \, K_{m:k}\, \frac{\Delta \rho(p_t:n;k\epsilon_x)}{\sqrt{\rho_{ref}(n;k\epsilon_x)}}.
\eea
Factors have been introduced so that the expression in square brackets is an average along the $k^{th}$ diagonal of normalized microbin covariances, with weighting factor ${\sqrt{\bar n_a \bar n_b}}/{\bar n_k} \sim 1$. The fourth line identifies the square bracket as a ratio of autocorrelation histograms. Factor $\bar n_k / \bar n \sim 1$ has been absorbed into the definition of the autocorrelation ratio, but could be extracted as a correction factor. In the fifth line the binning system has been shifted by 1/2 bin according to Fig.~\ref{intderiv} (third panel), and symmetry about the origin has been invoked (requiring an additional factor 2) to simplify the indexing. The histograms are finally converted to densities by including  the microbin width. The last line, generalized to 2D on angle variables $(\eta,\phi)$, is Eq.~(\ref{inv}). That integral equation provides computationally cheap $O(n)$ access to autocorrelations.

\section{Inversion and regularization}

Eq.~(\ref{inv}), a discrete Fredholm integral equation, is a matrix equation of the form ${\bf D = T\, I + N}$, with data {\bf D}, image {\bf I} and statistical noise ${\bf N}$~\cite{invtheory}. In principle, one could simply invert the matrix equation to obtain the image. However, ${\bf T^{-1}}$ is effectively a differentiation and acts therefore as a high-pass filter in the language of electrical engineering. The ${\bf T^{-1}\, N}$ statistical noise term strongly dominates the image derived from a simple matrix inversion, and must be substantially reduced by smoothing or `regularization'~\cite{smooth}. The procedure involves treating the image as a matrix of free values in a $\chi^2$ fit subject to {\em Tikhonov regularization}: minimize $\chi^2_\alpha \equiv {\bf ||D - T\, I_\alpha||^2 + \alpha\, ||L\, I_\alpha||^2}$, including Lagrange multiplier $\alpha$ which controls the role of local gradient operator ${\bf L}$. The first term measures the data--integrated-image mismatch, the second term measures small-wavelength noise on the image. The latter is equivalent to a compensating low-pass filter which offsets the effect of the differentiation. The resulting image is represented by $\bf I_\alpha = T_\alpha^{-1}\, (D - N_\alpha)$ which estimates true image ${\bf I}$. 

\begin{figure}[h]
\begin{minipage}{19pc} \hfil
\includegraphics[width=9pc]{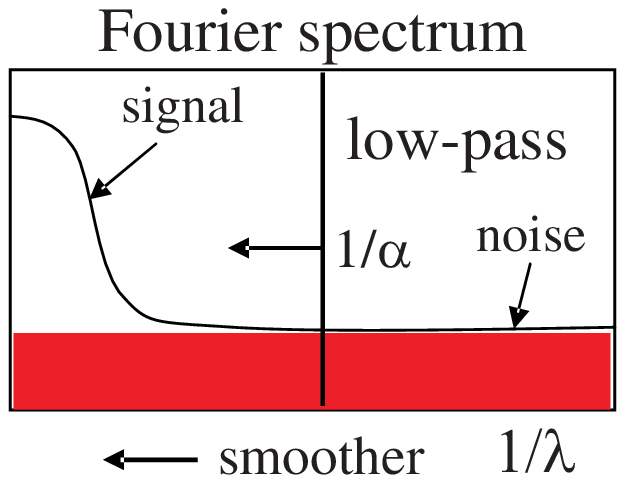} \hfil
\includegraphics[width=8pc]{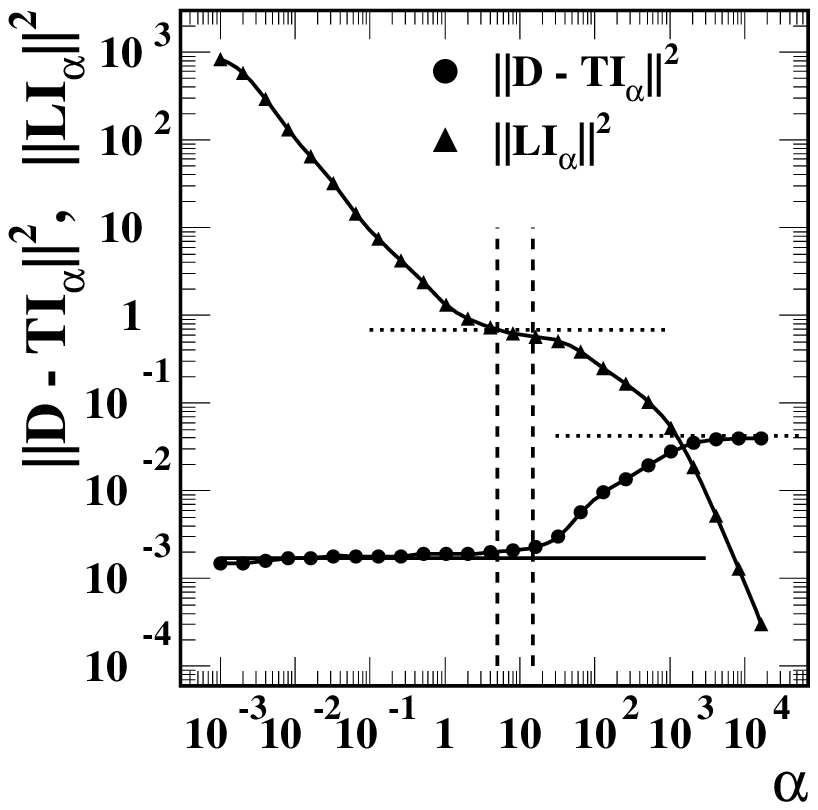} \hfil
\end{minipage} 
\begin{minipage}{19pc}
\includegraphics[width=9pc]{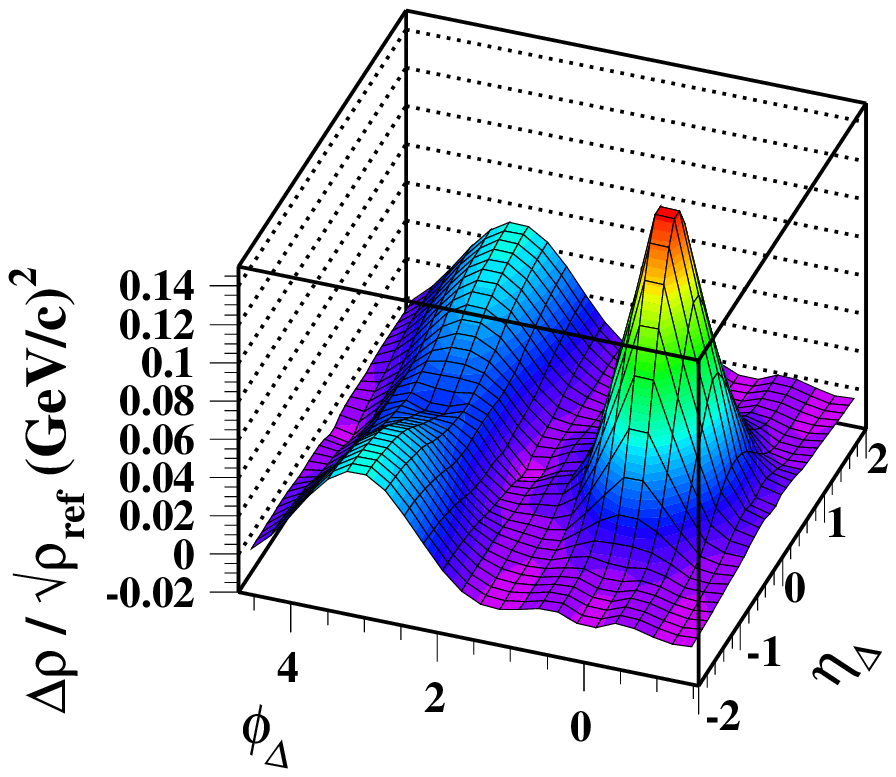} \hfil
\includegraphics[width=9pc]{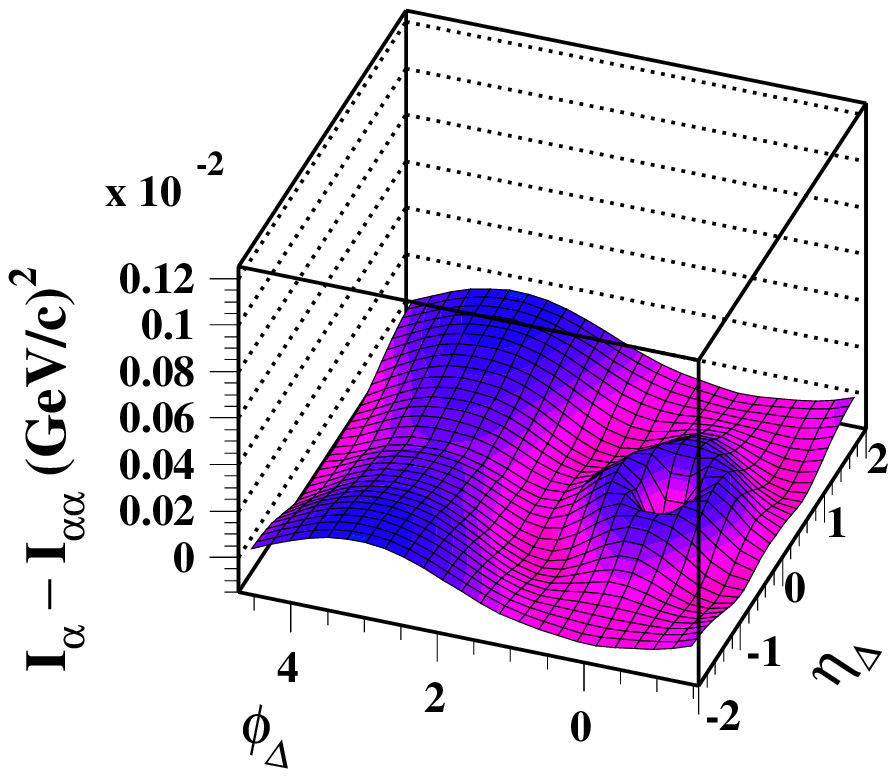} \hfil
\end{minipage} 
\caption{\label{inverse}  Sketch of Fourier power spectrum, distribution of two terms in $\chi^2$ on Lagrange multiplier $\alpha$, autocorrelation from fluctuation inversion and corresponding smoothing error.}
\end{figure}

Regularization is illustrated in a power-spectrum context by Fig.~\ref{inverse} (first panel): a tradeoff between information loss and noise suppression. The basis for choosing the optimum $\alpha$ is illustrated in the second panel. ${\bf ||D - T\, I_\alpha||^2} $ (dots) is signal loss and ${\bf ||L\, I_\alpha||^2}$ (triangles) is residual noise. As noted, $\alpha$ controls a compensating low-pass filter: small values retain all signal and a large amount of noise from the differentiation, larger values reduce noise, and finally distort the signal by over smoothing. The optimum value is determined as in Fig.~\ref{inverse} (second panel). This example illustrates  that there are clear criteria for choosing an optimum $\alpha$ so that negligible information is lost from the image while statistical noise is greatly attenuated.

Statistical and systematic errors are determined by looping through integration and differentiation twice in the sequence ${\bf D \rightarrow I_\alpha \rightarrow D_\alpha \rightarrow I_{\alpha \alpha}}$, including inversion, forward integration and second inversion. Difference ${\bf D - D_\alpha}$ estimates statistical error on the data and may itself be inverted to determine residual statistical error on image ${\bf I_\alpha}$. Difference ${\bf I_\alpha - I_{\alpha \alpha}}$ estimates the smoothing error. Fig.~\ref{inverse} (right two panels) shows image ${\bf I_\alpha}$ and corresponding smoothing error ${\bf I_\alpha - I_{\alpha \alpha}}$ for data from the Hijing Monte Carlo~\cite{hijing}, dominated by jet correlations~\cite{hijsca}.



\begin{figure}[h]
\begin{minipage}{19pc}
\includegraphics[width=9pc]{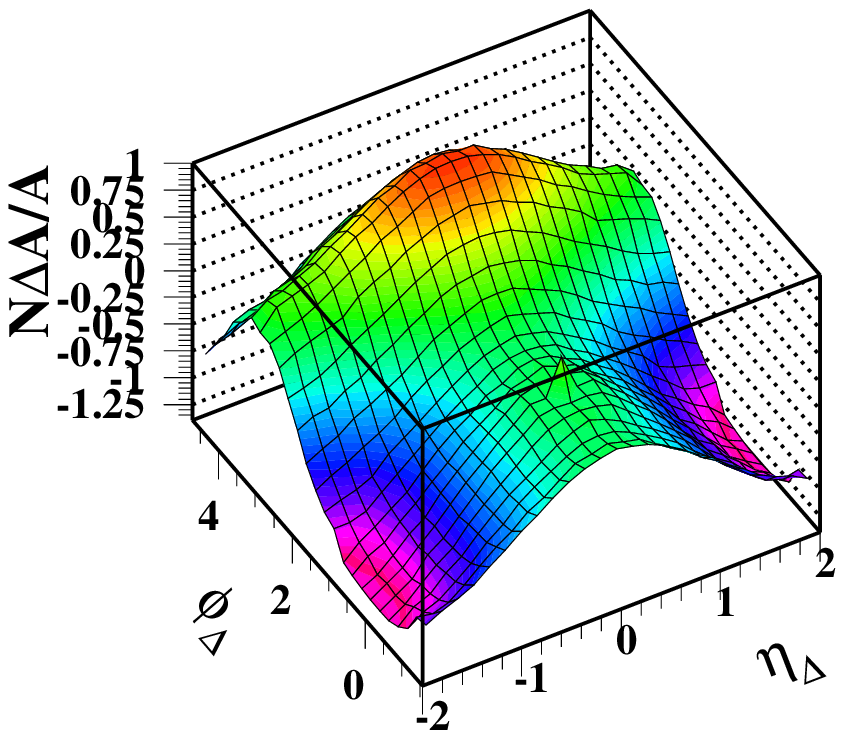}
\includegraphics[width=9pc]{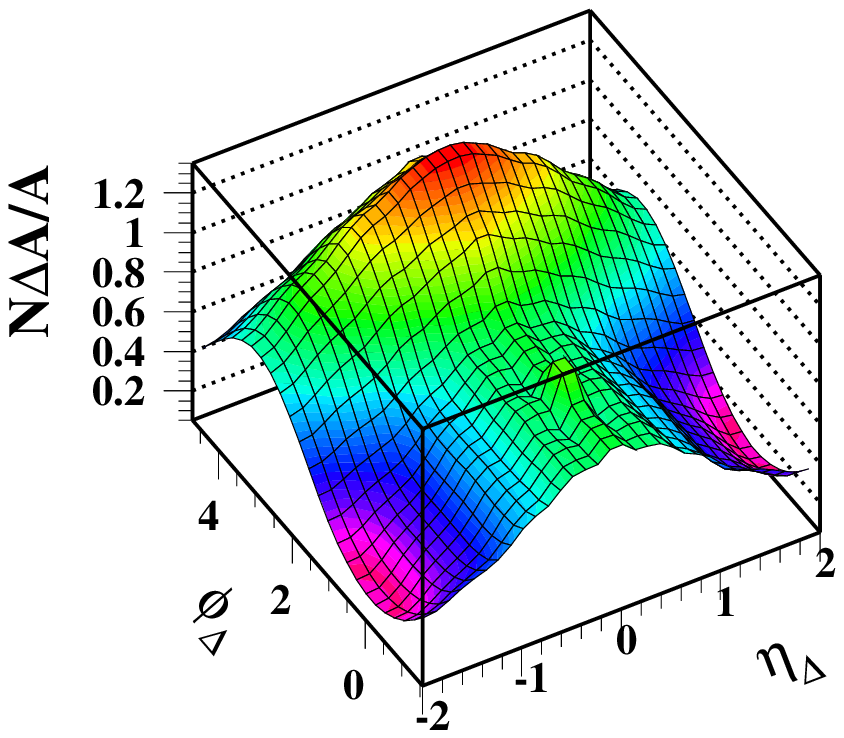}
\end{minipage} 
\begin{minipage}{19pc}
\includegraphics[width=9pc]{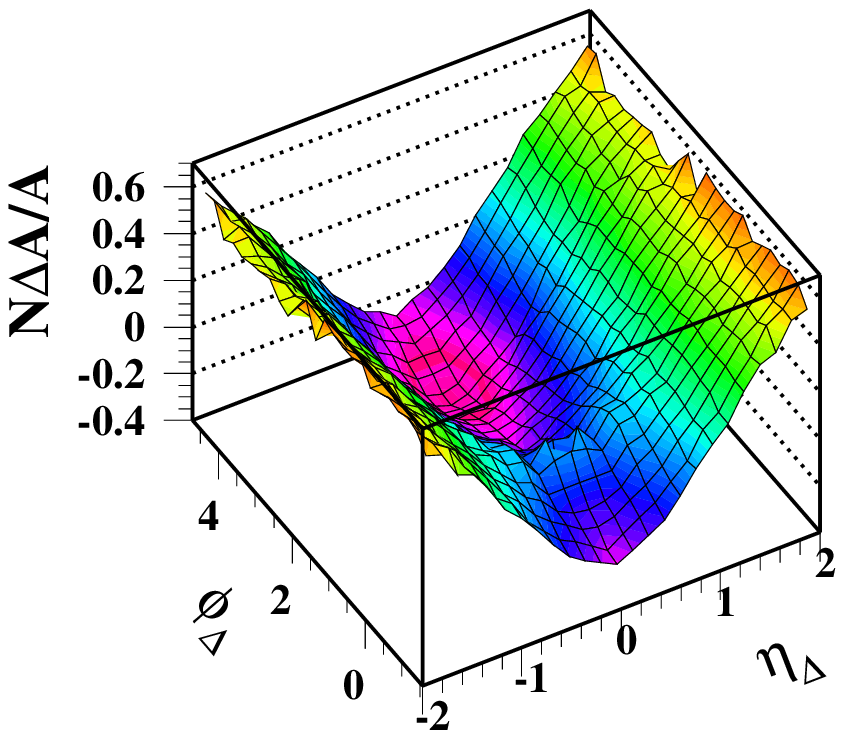}
\includegraphics[width=9pc]{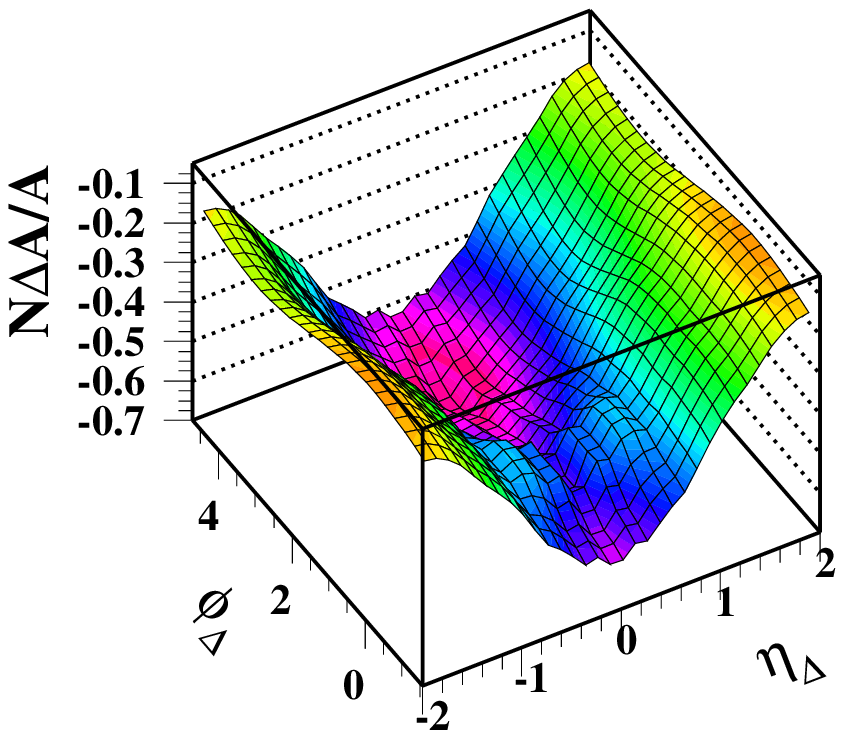}
\end{minipage} 
\caption{\label{precision}  Autocorrelations from the Pythia Monte Carlo for charge-independent (left panels) and charge-dependent (right panels) correlations and direct (left) {\em vs} inversion (right) methods.}
\end{figure}  

A precision comparison of autocorrelations obtained directly by pair counting and indirectly by fluctuation inversion is presented in Fig.~\ref{precision}. The data were obtained from the Pythia Monte Carlo~\cite{pythia}. The four panels in sequence are charge-independent direct and inverted autocorrelations and charge-dependent direct and inverted autocorrelations. The agreement between methods is excellent. 

\section{Correlation types}

\begin{figure}[h]
\begin{minipage}{38pc}
\includegraphics[width=9.5pc]{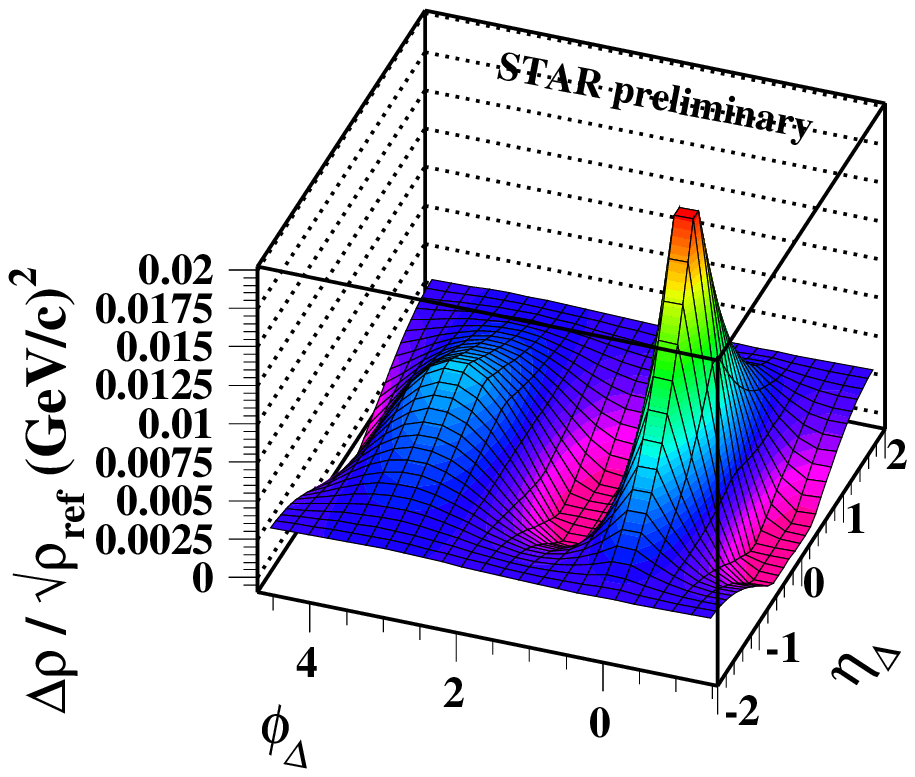}
\includegraphics[width=10.5pc]{sigdelspace}
\includegraphics[width=9pc]{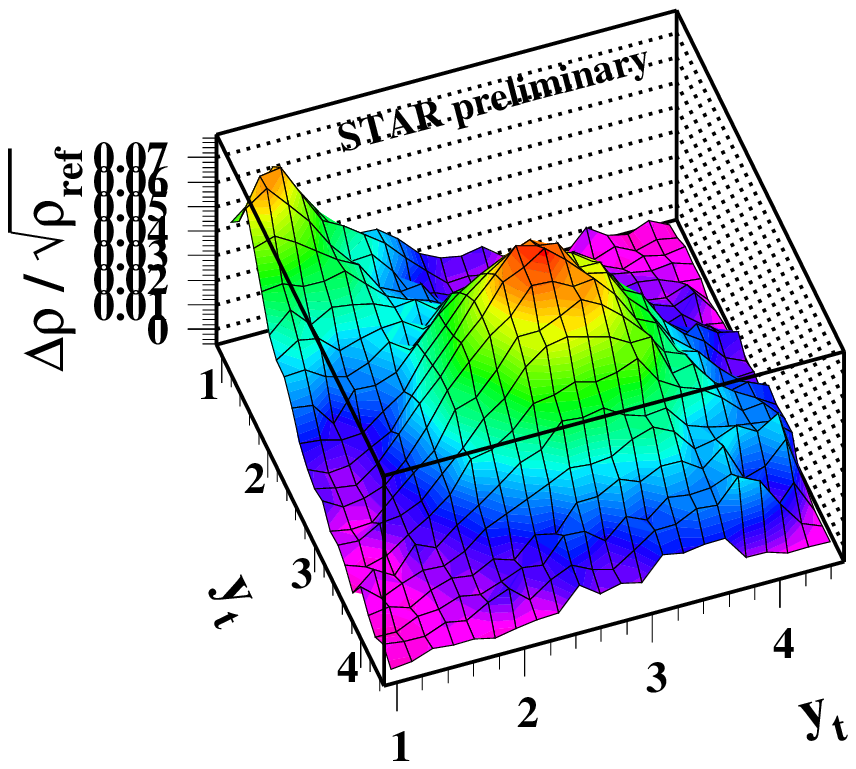}
\includegraphics[width=7.9pc]{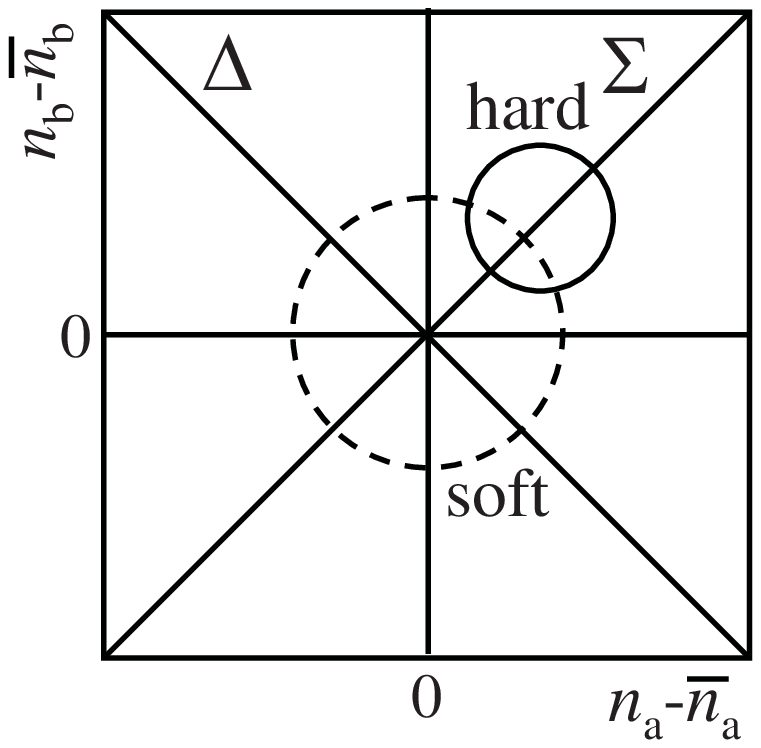}
\end{minipage} 
\caption{\label{corrtyp} $p_t$ autocorrelation with regions of positive and negative covariance, distributions of bin-pair contents illustrating corresponding correlation (solid) and anticorrelation (dash-dot) trends, distribution on $(y_{t1},y_{t2})$ illustrating soft and hard components, and distributions of bin-pair elements illustrating the role of rare hard events in producing positive-definite covariance.}
\end{figure}

Fig.~\ref{corrtyp} illustrates types of correlation measurements. The left panels show an autocorrelation obtained by $\langle p_t \rangle$ fluctuation inversion (first panel) and a sketch of corresponding two-bin correlations (second panel) for an arbitrary bin separation. 
In this type of correlation, structure may occur in most or all events, but with random position on $(\eta,\phi)$. The autocorrelation of a positive-definite measure (such as $n$ or $p_t$) must be positive-definite at the origin (a variance). Elsewhere (covariances), positive autocorrelation bins indicate correlation (solid curve in second panel), negative bins indicate {\em anti}correlation (dash-dot curve), as in the magenta areas adjacent to the positive same-side peak in the first panel. The dashed circle represents the reference.

The right panels illustrate a case where localized structure with approximately fixed position is present in a minority of events. The third panel shows correlations on transverse rapidity $y_t$ (that distribution is {\em not} an autocorrelation). In the fourth panel distributions are sketched for two event classes. The dashed curve represents common ({\em e.g.,} soft) events, the solid circle represents exceptional events ({\em e.g.,} hard events which contain a detectable parton scatter) from  an ensemble of p-p collisions. Positive covariances result from the exceptional events which have an additional multiplicity contribution localized on the $y_t$ space and (in contrast to the autocorrelation example at left) occuring at a nearly fixed position over the event ensemble. The normalized covariance density comparing sibling and mixed pairs in the third panel reveals the contribution from exceptional events (parton fragments).


\section{Autocorrelations and conditional distributions}

\begin{figure}[h]
\begin{minipage}{38pc}
\includegraphics[width=18.9pc]{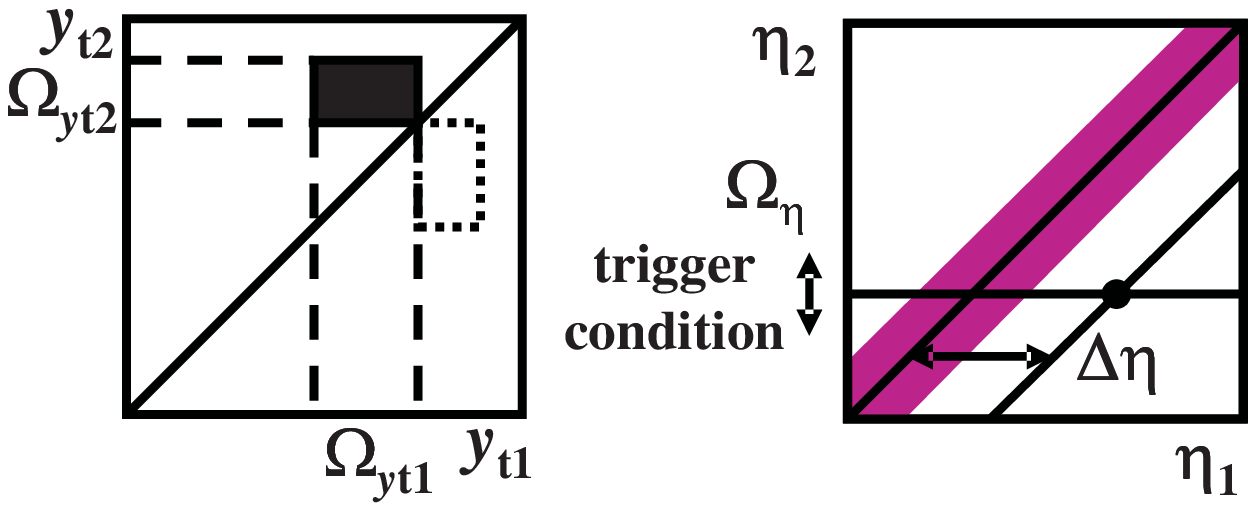}
\includegraphics[width=18pc]{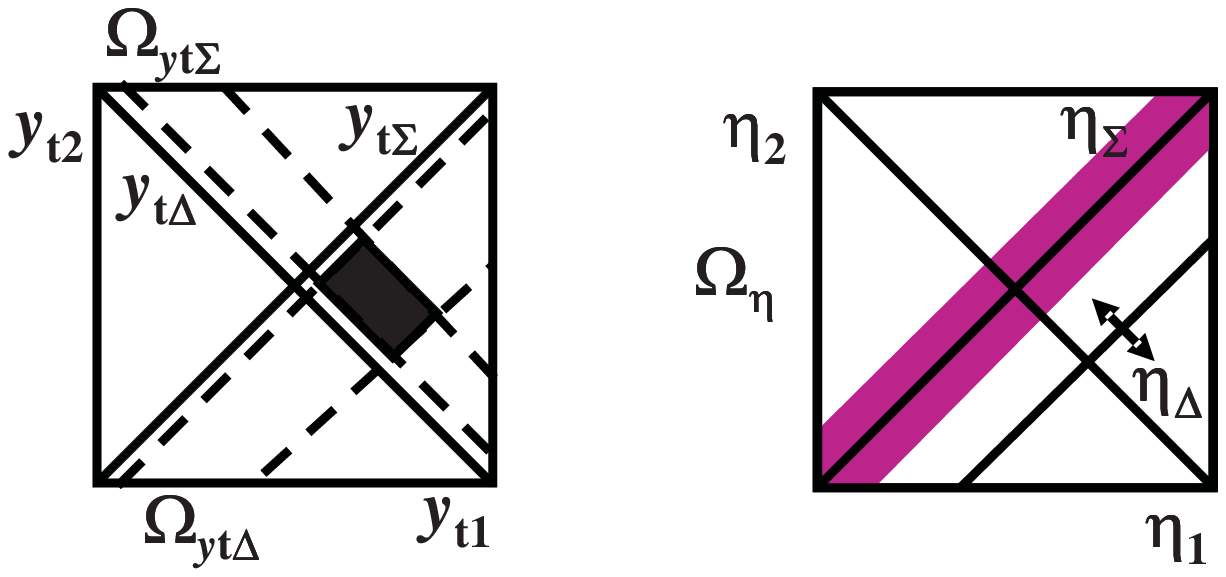}
\end{minipage} 
\caption{\label{autocond}   Conditions defined on $(y_{t1},y_{t2})$ and $(\eta_1,\eta_2)$ for a leading-particle (trigger) analysis compared to a condition on $(y_{t\Sigma},y_{t\Delta})$ and autocorrelation on $\eta_\Delta$ invoking no leading particle.}
\end{figure}

Conventional studies of jet correlations in A-A collisions employ a {\em leading-particle} analysis (invoked when full jet reconstruction is not possible)~\cite{leading}. The goal is to estimate a parton momentum by that of the highest-$p_t$ (above some threshold) particle in a collision---the leading or trigger particle. The analysis utilizes two or three {\em conditional} distributions as illustrated in Fig.~\ref{autocond} (first two panels), where for the sake of comparison transverse momentum $p_t$ has been replaced by transverse rapidity $y_t$. The condition on $(y_{t1},y_{t2})$ is a rectangle representing asymmetric trigger- and associated-particle $y_t$ conditions.  Trigger region $(\Omega_{t_{t1}},\Omega_{t_{t2}})$ is by construction displaced from the sum diagonal. Angular correlations are also defined as conditional distributions relative to the trigger-particle position in single-particle angle space $(\eta,\phi)$. The angular correlation is plotted (using pseudorapidity as an example) on {\em conditional} angle difference $\Delta \eta \equiv \eta - \eta_{trigger}$, which represents an event-wise shift of the {\em single-particle} angle origin. 

Alternatively, one can abandon attempts to estimate parton $p_t$ {\em per se} and use a technique involving no conditions on momentum, or a symmetric condition on $(y_{t1},y_{t2})$ with no conditions on angle variables, as illustrated in Fig.~\ref{autocond} (second two panels). The fragment distribution on $(y_{t1},y_{t2})$ can be studied in its own right. Cut conditions on $(y_{t1},y_{t2})$ can be defined to study corresponding changes in jet angular morphology on ($\eta,\phi$). Angular autocorrelations invoking no trigger condition and defined on symmetric difference variables such as $\eta_\Delta \equiv \eta_1 - \eta_2$ access a minimum-bias parton population. Difference variable $\eta_\Delta$ spans the diagonal axis of the 2D $(\eta_1,\eta_2)$ space. Variables $\Delta \eta$ and $\eta_\Delta$ are numerically different, support different distributions and should not be confused (the same comment applies to $\Delta \phi$ and $\phi_\Delta$).
The correlation measures used in the two cases may be compared:
\bea
\frac{\Delta \rho(n;\eta_\Delta,\phi_\Delta)}{\sqrt{\rho_{ref}(n;\eta_\Delta,\phi_\Delta)}} \hspace{.2in} vs \hspace{.2in} \frac{1}{N_{trig}}\frac{d^2N_{pair}}{d\Delta \eta\, d\Delta \phi} \equiv \frac{\rho(n;\Delta \eta,\Delta \phi)}{\int_{\Omega_{trig}}dp_t\, \rho(n;p_{t,trig})}.
\eea
The latter depends directly on leading- or trigger-particle acceptance $\Omega_{trig}$, and a reference must be provided {\em a posteriori} by defining a model function to describe the background.

\section{Summary}

We have derived an integral equation which connects fluctuation scale dependence to corresponding autocorrelations. Inversion of the integral equation reveals autocorrelations which are equivalent to those from pair counting. Fluctuations are thereby interpretable in terms of underling two-particle correlations. Autocorrelations are complementary to leading-particle techniques for analysis of jet correlations. Definition of normalized variances, covariances and variance differences is tightly constrained by a number of considerations, including linearity, relating fluctuations to correlations, minimizing statistical bias and insuring that correlation structure is maximally accessible and interpretable over a broad range of contexts.

This work was supported in part by the Office of Science of the U.S. DoE under grant DE-FG03-97ER41020.

\end{document}